\documentclass[onecolumn,pre]{revtex4-1}

\usepackage{graphicx}
\usepackage{pstricks}
\usepackage{amsmath}

\newcommand{\be}{\begin{equation}}
\newcommand{\ee}{\end{equation}}
\newcommand{\ben}{\begin{displaymath}}
\newcommand{\een}{\end{displaymath}}
\newcommand{\bea}{\begin{eqnarray}}
\newcommand{\eea}{\end{eqnarray}}
\newcommand{\bean}{\begin{eqnarray*}}
\newcommand{\eean}{\end{eqnarray*}}

\newcommand{\ind}[1]{_{\rm #1}}

\newcommand{\Eq}[1]{Eq.~(\ref{#1})}
\newcommand{\Fig}[1]{Fig.~\ref{#1}}

\newcommand{\nn}{\nonumber}
\newcommand{\ep}{\varepsilon}
\newcommand{\brac}[1]{\left( #1 \right)}
\newcommand{\Brac}[1]{\left[ #1 \right]}
\newcommand{\BRAC}[1]{\{ #1 \}}
\newcommand{\mean}[1]{\langle #1 \rangle}

\newcommand{\derive}[2]{\frac{\partial \, #1}{\partial \, #2}}
\newcommand{\dderive}[2]{\frac{\partial^2 \, #1}{\partial \, #2^2}}

\newcommand{\vecr}{\vec r}

\begin{document}
 
\title{Spatially Resolved Dynamic Structure Factor of Finite Systems from MD Simulations}

\author{Thomas Raitza}
\email{thomas.raitza@uni-rostock.de}
\affiliation{Institut f\"ur Physik, Universit\"at Rostock,\\
  18051 Rostock, Germany}

\author{Heidi Reinholz}
\email{heidi.reinholz@jku.at}
\affiliation{Institut f\"ur Theoretische Physik, Johannes-Kepler- 
Universit\"at Linz,\\
  4040 Linz, Austria}
\affiliation{Institute of Physics, University of Western Australia,\\  
Perth, 6009 WA, Australia}

\author{Igor Morozov}
\email{morozov@ihed.ras.ru}
\affiliation{Joint Institute for High Temperatures of RAS,\\
Izhorskaya, 13, build.~2,\\
Moscow 125412, Russia}

\author{Gerd R\"opke}
\email{gerd.roepke@uni-rostock.de}
\affiliation{Institut f\"ur Physik, Universit\"at Rostock,\\
  18051 Rostock, Germany}

\date{\today}

\begin{abstract}
The dynamical response of metallic clusters up to $10^3$ atoms is investigated using the restricted molecular dynamics simulations scheme. Exemplarily, sodium like material is considered. Correlation functions are evaluated to investigate the spatial structure of collective electron excitations and optical response of laser excited clusters. In particular, the spectrum of bi-local correlation functions shows resonances representing different modes of collective excitations inside the nano plasma. The spatial structure, the resonance energy and width of the eigenmodes have been investigated for various values of electron density, temperature, cluster size and ionization degree. Comparison with bulk properties is performed and the dispersion relation of collective excitations is discussed.
\end{abstract}

\maketitle

\section{Introduction}

Nano plasmas can now be readily produced in laser irradiated clusters, and new physical phenomena have come into focus experimentally as well as theoretically. Interactions between laser fields of $10^{13} - 10^{16}$ W cm$^{-2}$ and clusters have been investigated over the last few years, see Refs. \cite{Mac94a}--\cite{Doeppner07}. After laser interaction, extremely large absorption rates of nearly 100\%, see \cite{Dit97b}, as well as x-ray radiation, see Refs. \cite{Mac94b}--\cite{Mic07}, were found. In pump-probe experiments, e.g.\, by D\"oppner {\it et al.} \cite{Doeppner07} and Fennel {\it et al.} \cite{Fennel07}, the absorption rate of a second laser pulse is strongly dependent on the time delay what is caused by the dynamical properties of the expanding cluster. We will discuss the dynamical response function of the electrons in a nano plasma that is responsible for scattering and absorption of electromagnetic radiation.

Collective electronic excitations of the nano plasma usually are interpreted as Mie resonances of a homogeneously charged sphere. Absorption cross section experiments by Xia {\it et al.} \cite{Kresin09} show multiple resonance structures indeed. The effect of collective electron motion can also be seen in ultraviolet (UPS) and x-ray photoelectron spectroscopy (XPS) experiments, see \cite{Huefner},  which are used to detect binding energies of core level electrons in small metal clusters \cite{Senz09,Andersson11}. In fusion related experiments by Ditmire {\it et al.} \cite{Dit99}, Grillon {\it et al.} \cite{Gri02}, as well as Madison {\it et al.} \cite{Madison} ignition processes are started via irradiation of deuterium clusters. Collective effects in the optical response are discussed in the context  of metallic nanoshells by H\"oflich {\it et al.} \cite{Hoefflich09} as well as nanocavities by Maier {\it et al.} \cite{Verellen09}.

In theoretical  calculations of finite systems, see Raitza {\it et al.} \cite{CPP09,Raitza09,Raitza10,RaitzaPHD}, a more complex resonance structure was found. Earlier investigations by Reinhard {\it et al.} \cite{Reinhard96} and Kull {\it et al.} \cite{Greschik04,Arndt06} led to comparable results. The method of resonance structure analysis using spherical harmonics is known from the discussion of giant dipole resonances of nuclei, see Reinhard {\it et al.} \cite{Reinhard07}. Quantum and semi-classical methods, see Refs. \cite{Cal00}--\cite{Saalmann08}, respectively,  were used to investigate the cluster excitation via laser fields. Collisional absorption processes in nano plasmas have been the subject of theoretical investigations by Hilse {\it et al.} \cite{Hilse05}. Using density functional theory (DFT) calculations, the electronic structures of cold clusters were analyzed by Ekardt \cite{Ekardt84}, K\"ummel {\it et al.} \cite{Kuemmel2000}, Brack {\it et al.} \cite{Brack08}, as well as Krotscheck {\it et al.} \cite{Krotscheck05}. The damping of collective electron oscillations was investigated by Ramunno {\it et al.} \cite{Brabec06} emphasizing the importance of collisional processes beside the Landau damping.

In this work, molecular dynamics (MD) simulations will be used to study nano plasmas in metal clusters. Clusters consisting of 55 up to 1000 sodium like atoms are considered after short pulse laser irradiation with intensities in the order of $10^{12}$ Wcm$^{-2}$. Properties of the nano plasma are mainly determined by the dynamics of electrons which are bound to the cluster but ionized from the former atoms, comparable to conduction electrons in bulk systems. As already shown in earlier publications, see \cite{CPP09}, plasma parameters as known from bulk (temperature and particle density) but also the cluster size and net charge are justified for characterization since the electrons can be assumed to be in local thermal equilibrium within time scales considered here. We focus on parameter ranges where the plasma can be treated classically. Strong correlations are taken into account via collisions of all particles. Concepts that have been well established for infinite bulk systems near thermodynamic equilibrium  have to be modified for applications to finite systems, e.g. clusters. In particular, we are interested in the dynamical structure factor and the response function for such finite nano plasmas.  In order to bridge from finite systems to bulk plasmas, we investigate size effects, e.g. in the dynamical collision frequency. First results in this direction have been reported in Refs. \cite{CPP09,CMT30,CMT31}.

In Sec. II,  correlation functions and their relation to optical properties of homogeneous bulk plasma  are introduced as far as it will be of interest to extend the approach to finite systems. Expressions will also be used for   comparison with nano plasmas  in the limit of large clusters. Sec. III explains the restricted molecular dynamics (RMD) scheme for the calculation of the particle trajectories from which the total and bi-local current density correlation functions are determined. Symmetries in the correlation matrix  discussed in Sec. IV can be used for an improved statistics. The decomposition of the correlation matrix into eigenvectors and eigenvalues is interpreted as a decomposition into collective excitation modes. In Sec. V, the excitation modes will be characterized with respect to spherical harmonics. In the further analysis, we focus on modes with a dipole moment, which are also seen in the total current density auto-correlation function. First results for resonance frequencies and damping are presented. Regarding the dipole-like modes, the spatial structure at the selected resonance frequency will be discussed in Subsec. A and Subsec. B. Conclusion and outlook are given in Sec. VI.

\section{Linear response theory of plasmas in equilibrium}

Within linear response theory as derived by Kubo {\it et al.}, see \cite{Kubo,Zubarev}, the reaction of a many-particle system to weak external perturbations can be related to the dynamical behavior of fluctuations in thermal equilibrium. Denoting the equilibrium statistical operator with $\rho\ind{0}$, we introduce the two-time correlation function of the fluctuations $\delta A_i (t) = A_i (t) - {\rm Tr} \BRAC{A_i \rho\ind{0}}$ as the Kubo scalar product
\be
\brac{A_i (t); A_j (0)} = \beta \int_0^1 {\rm d} \lambda \, {\rm Tr} \BRAC{\delta A_i (t) \delta A_j ({\rm i} \hbar \beta \lambda) \rho\ind{0}},
\ee
where the time dependence is given in the Heisenberg picture. The indices $i$ and $j$ identify quantum observables. In particular we consider local properties so that they contain also the position $\vec r$. In case of $i = j$, it is called auto-correlation function (ACF).

In the classical case, equilibrium two-time correlation functions can be calculated according to
\be
\brac{A_i (\vec r, t); A_j (\vec r', 0)} = \lim_{T \to \infty} \frac{1}{T} \int_0^{T} \, {\rm d} \tau \, \delta A_i (\vec r, \tau + t) \cdot \delta A_j (\vec r', \tau),\label{correlklass}
\ee
where we assumed ergodic systems - the ensemble average can be replaced by a time average. The spectrum of the equilibrium correlation function $\mean{A_i (\vec r); A_j (\vec r')}_{\omega}$ then results from Laplace transformation.

We consider an induced electron density fluctuation $\delta n\ind{e} (\vec r, t) = n\ind{e} (\vec r, t) - n\ind{e,0} (\vecr)$  at time $t$ as the deviation from the equilibrium density distribution $n\ind{e,0} (\vecr)$ due to an external potential $U\ind{ext} (\vec r', t')$ at times $t' < t$. Close to equilibrium, the correlation between the external potential and the induced density fluctuation is only dependent on the time difference $\Delta t = t - t'$. Thus, one is able to discuss its spectrum after Laplace transform. In the same way, the induced electrical current density $\vec j\ind{e} (\vec r, t)$ is related to the external electric field $\vec E (\vec r', t')$. Via Kubo's theory, these induced quantities, $\delta \mean{n\ind{e} (\vec r)}_\omega$ and $\delta \mean{\vec j (\vec r)}_\omega$, can be expressed within linear response, see \cite{Zubarev}, as
\bea \label{korel_nnp}
\delta \mean{n\ind{e} (\vec r)}_\omega & = & \beta \int {\rm d}^3 \vec r' \, \mean{\delta n\ind{e} (\vec r); \delta \dot{n}\ind{e} (\vec r')}_\omega \, U\ind{ext} (\vec r', \omega),\\
\delta \mean{\vec j (\vec r)}_\omega & = & \beta \int {\rm d}^3 \vec r' \, \mean{\vec j (\vec r); \vec j (\vec r')}_\omega \, \vec E (\vec r', \omega).
\eea
The spectrum of the density fluctuation correlation $\mean{\delta n\ind{e} (\vec r); \delta \dot{n}\ind{e} (\vec r')}_\omega$ is related to a scalar response function. The current-density correlation $\mean{\vec j (\vec r); \vec j (\vec r')}_\omega$ represents in general a tensor due to the directions of the current density vector.

Before considering non-local response functions, we shortly mention homogeneous systems. Properties of the bulk plasmas with electron density $n\ind{e}$ and inverse temperature $\beta$ are only dependent on the difference of the positions $\Delta \vec r = \vec r - \vec r'$. Thus, after Fourier transform of the spatial difference $\Delta \vec r$, the correlations are dependent on a wave vector $\vec k$.

The dynamical structure factor is directly related to the density fluctuation correlation, as
\be \label{strfac}
S (\vec k, \omega) = \frac{1}{2 \pi N} \mean{\delta n_k; \delta n_k}_{\omega}
\ee                                                                                          
with $N$ the number of particles. For  further relations to the dielectric function and the optical response of a homogeneous bulk plasma see \cite{ReinholzHabil}. Note that the density fluctuations \Eq{korel_nnp} as well as the density correlation function in \Eq{strfac} can be expressed in terms of the current-density correlation function via partial integration and using the continuity equation. Thus, the dynamical structure factor is divided into a static part $S\ind{0} (\vec k)$ and a dynamical part which is directly related to the longitudinal part of the current-density correlation function
\be
S (\vec k, \omega)  = \frac{S\ind{0} (\vec k)}{- {\rm i} \omega} + \frac{1}{2 \pi N} \frac{k^2}{\omega^2} \mean{\vec j^{||}_k; \vec j^{||}_k}_{\omega}.
\ee
It is of fundamental interest to describe the collective behavior of the system as response to external fields, in particular emission, absorption and scattering of light. In bulk systems, the wave vector and frequency dependent response function reads
\be
\chi(\vec k, \omega)= -{\rm i} \beta \Omega\ind{0} \frac{k^2}{\omega} \mean{\vec j^{||}_k;  \vec j^{||}_k}_{\omega},
\ee
which can be evaluated using quantum statistical approaches such as Green function theory, see \cite{ReinholzHabil}, or numerical approaches such as MD simulations, see \cite{Morozov05}. As collisions are relevant in strongly correlated systems, the dynamical collision frequency $\nu(\omega)$ is derived and appears in a  generalized Drude formula  \cite{Reinholz00,Reinholz04}
\be \label{gendrude}
\lim_{k\rightarrow 0} \chi(\vec k, \omega)=\;
\frac{\ep\ind{0} k^2\, \omega_{\rm pl}^2}{\left(\omega^2 - \omega_{\rm pl}^2\right) + {\rm i}\omega \nu(\omega) }.
\ee
In the classical case, the current-density correlation function has been extensively discussed in the long wavelength limit $k \to 0$ applying MD simulations and perturbative approaches. Exemplarily, we refer to \cite{Morozov05}.

The state of a homogeneous one-component plasma in thermodynamic equilibrium is characterized by the nonideality parameter $\Gamma = e^2 (4 \pi n\ind{e} / 3)^{1/3} (4 \pi \ep\ind{0} k\ind{B} T\ind{e})^{-1}$ and the degeneracy parameter $\Theta = 2 m\ind{e} k\ind{B} T\ind{e} \hbar^{-2} (3 \pi^2 n\ind{e})^{-2/3}$, $T\ind{e}$ is the temperature of the electrons. Considering the response function $\chi(\vec k, \omega)$ in the long wavelength limit, a sharp peak arises at the plasmon frequency $\omega_{\rm pl}$, see \Eq{gendrude}. For finite wavelengths, the resonance is shifted and can be approximated by the so called Gross-Bohm plasmon dispersion for small wave numbers $k$, see \cite{GrBo,KKER86}, $\omega(k) \approx \omega_{\rm pl} + 3 k^2/\kappa^2 + ...$  with the Debye screening length $\kappa^{-1} = [n\ind{e} e^2/(\epsilon_0 k_B T\ind{e})]^{-1/2}$. This relation has recently been revisited with respect to the relevance of collisions by Thiele {\it et al.} \cite{Thiele}. According to \Eq{gendrude}, the general behavior of the response function $ 
\chi(\vec k, \omega)$ in the long-wavelength limit is closely related to the collision frequency which is relevant in non-ideal plasmas, see \cite{Thiele,Reinholz04}. In the two-component plasma, a phonon mode can arise in addition to the plasmon excitations \cite{Morozov98}.

The response function $\chi(\vec k, \omega)$ and the related dynamical
structure factor  $S(\vec k, \omega)$ as well as the optical properties
have been intensively investigated for electron-ion bulk systems, see  
Refs.
\cite{Thiele, Fortmann09}. 
In this work, the inhomogeneous case of finite clusters in local
thermal equilibrium will be discussed. The response of inhomogeneous
systems is not only dependent on the difference of the positions, but on
$\vec r$ and $\vec r'$ separately. Therefore, spatially resolved current
density correlation functions $\mean{\vec j (\vec r); \vec j (\vec
r')}_{\omega}$  can not be diagonalized by spatial Fourier transform.
Instead of plane waves, other basis functions have to be found in  
order to
characterize the  collective excitations of electrons.

\section{MD simulations of finite plasmas}

Finite plasma systems have  been investigated using the restricted molecular dynamics (RMD) simulations, see Raitza {\it et al.} \cite{CPP09}. A two-component system of singly charged ions and electrons will be described using an error function pseudo potential for the interaction of particles $i$ and $j$
\be
V\ind{erf} (r_{ij}) = \frac{Z_{i} Z_{j} e^2}{4 \pi \ep\ind{0} r_{ij}} {\rm erf} \brac{\frac{r_{ij}}{\lambda}},
\label{erf}
\ee
where $Z_{i}$ is the charge of the $i$th particle. The Coulomb interaction is modified at short distances, assuming a Gaussian wave function for electrons  motivated by the account of quantum effects. Considering a sodium like system, the potential parameter $\lambda = 0.318$ nm was chosen in order to reproduce the ionization energy of $I\ind{P} = V\ind{ei} (r \rightarrow 0) = -5.1$~eV for solid sodium, as already discussed for MD simulations by Suraud \textit{et al.} \cite{Suraud06}.

The velocity Verlet algorithm \cite{Verlet67} was applied to solve the classical equations of motion for electrons and ions. This method takes into account the conservation of the total energy of the finite system, as long as there is no external potential. To follow the fast electron dynamics, time steps of $0.01$ fs were taken to calculate the time evolution. Contrary to bulk MD simulations no periodic boundary conditions are applied.

Icosahedral arrangements of 55, 147, and 309 ions, see \cite{CPP09}, were considered as initial configuration for the ion positions. For these nearly spherically, homogeneously distributed ions, the ion density typical for solid sodium is given by an ionic next neighbor distance of $d\ind{0} = 0.212$ nm. In addition, randomly distributed ion configurations within a given sphere were considered for comparison and the number of ions was increased up to 1000 particles. Starting with a neutral cluster, the electrons have been positioned nearby the ions with small, randomly distributed deviations from the ion positions.

To simulate experiments where clusters are excited by short pulse lasers, MD simulations are performed under the influence of an electric field, assuming a Gaussian shape and pulse duration of about 100 fs. Due to the largely increased kinetic energy of the electrons, ionization processes occur. After the laser field is switched off, the ionization degree of the cluster is determined by the number of electrons found outside the cluster radius with positive total energy, so that they can escape from the cluster. Due to ion excitation on larger time scales, a slow expansion of the positively charged cluster is obseved \cite{Jortner}, leading to Coulomb explosion experimentally.

Considering the single-time properties, it was found in \cite{CPP09} that already local thermodynamic equilibrium (LTE) is established within a few fs  after the electron heating. In particular, at each time step, the momentum distribution of electrons is well described by a Maxwell distribution, and the spatial density profile agrees with a Boltzmann distribution with respect to the average potential that is determined by the actual ion configuration and the self-consistent electronic mean field. The fact that electrons are considered within  sub fs time intervals, while the ion configuration remains nearly unchanged, enables us to separate the electron dynamics from the ion dynamics.

Subsequently, the dynamical properties of the electron subsystem can be calculated for a frozen ionic configuration thus referring to a specific time. This is considered as an adiabatic approximation to the true dynamical properties of the electron subsystem which have to take into account the slow change in the ion configuration. More rigorously, non-stationary time dependent correlation functions have to be treated for the full charged particle system.

Using the RMD simulations scheme as introduced in \cite{CPP09}, the ions are kept fixed acting as external trap potential. 
Starting from an initial state, the many-electron trajectory $\{\vec r_l(t), \vec p_l(t)\}$ is calculated, solving the classical equations of motion of the electrons. From this, all further  physical properties of the electron subsystem inside the cluster are determined. 
Within RMD simulations, we consider no temporal variation of the plasma parameters that are determined by the frozen ion distribution, the electron temperature and the degree of ionization. A long-time run can be performed in order to replace the ensemble average by a temporal average. This has been successfully done for the single-time properties such as the momentum distribution and the density profile, see \cite{CPP09} and will now be applied to the two-time correlation functions.

Using classical MD simulation techniques, the results are valid for non-degenerate plasmas. This restricts the temperature range to $T \ge 1$ eV where our simulations can be compared with realistic sodium clusters. Values for the plasma parameter $\Gamma > 1$ can be treated since we are not confined to the weak coupling limit as, e.g., in perturbation theory.

\begin{figure}
\begin{center}
\includegraphics*[width=0.7\textwidth]{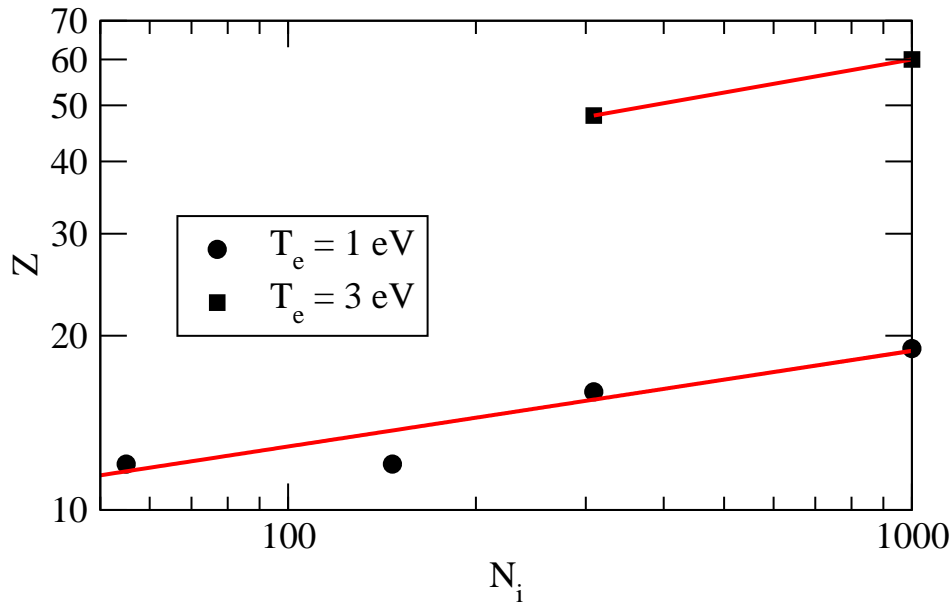}
\end{center}
\caption{(Color online) Simulation data of the cluster charge $Z$ (symbols) and a power fit (solid line) for clusters with different numbers of ions $N\ind{i}$ at different temperatures  is shown.}
\label{zfig}
\end{figure}

In our RMD calculations, we start from a homogeneous ion configuration (icosahedral or randomly distributed) inside the cluster at fixed ion density. In the case of random distribution, we perform averaging over different initial configurations of ions. The Langevin thermostat was used to heat the electrons at an initial stage. We have chosen the Langevin thermostat introducing a friction term with suitable sign to adjust the intended kinetic energy. Furthermore, a random source term is applied that thermalizes the system. Hot electrons are emitted during this stage so that the cluster becomes ionized. Evaluating the trajectories of electrons, sufficient time of about 200 fs has to be allowed before a stationary ionization degree is established. Then the thermostat is switched off and a data taking is performed using an ensemble at fixed number of density, volume and energy. It is checked that the mean cluster charge $Z$ and the system temperature do not change any more.  In \Fig{zfig}, the  cluster charge $Z$  depending on cluster size $N\ind{i}$ is shown. A power fit $Z (N\ind{i}) = A N\ind{i}^B$ with, for example,  $A = 0.165$ and $B = 0.197$ for $T\ind{e}=$1 eV shows the trend of the size dependent ionization degree.

Using the trajectories of all $N\ind{e}$ electrons obtained from the RMD simulations scheme, the local current density $\vec j\ind{e} (\vec r, t)$ at position $\vec r$ was calculated for each time step $t$
\be
\vec j\ind{e} (\vec r, t) = \lim_{\Delta V_{\vec r} \to 0} \, \frac{e}{m\ind{e}} \frac{1}{\Delta V_{\vec r}} \sum_{l = 1}^{N\ind{e}} \vec p_l (t) \, \delta_{\Delta V_{\vec r}} (\vec r_l (t)),
\label{momdens}
\ee
which is the sum over all electron momenta $\vec p_l$ inside a small volume $\Delta V_{\vec r}$ at position $\vec r$, where $\delta_{\Delta V_{\vec r}} (\vec r_l (t)) = 1$, and $\delta_{\Delta V_{\vec r}} (\vec r_l (t)) = 0$ for electrons found outside $\Delta V_{\vec r}$. The size of the volume determines the spatial resolution of the local current density $\vec j\ind{e} (\vec r, t)$. However, it must be taken sufficiently large to reduce statistical fluctuations.

The bi-local correlation tensor of the normalized spatially-resolved current density is calculated according to \Eq{correlklass} as
\be
\brac{\vec j\ind{e} (\vec r, 0); \vec j\ind{e}  (\vec r', t)} = \frac{\sum_{i = 1}^{N\ind{\tau}} \vec j\ind{e} (\vec r, i \cdot \tau) \otimes \vec j\ind{e} (\vec r', i \cdot \tau + t)}{N\ind{\tau} \cdot \mean{\vec j\ind{e} ^2}},\label{mdcorrt}
\ee
with $\vec j\ind{e}$ the total current density.   Typical values are  $N\ind{\tau} = 10^5-10^6$ and $\tau \sim 0.1$ fs. Its Laplace transform reads
\be
\mean{\vec j\ind{e} (\vec r); \vec j\ind{e} (\vec r')}_{\omega} = \int_0^{\infty} \, {\rm d} t \, {\rm e}^{{\rm i} \omega t} \, \brac{\vec j\ind{e} (\vec r, 0); \vec j\ind{e} (\vec r', t)}.\label{mdcorro}
\ee
In the following, we restrict ourselves to the diagonal components $\mean{j\ind{e}^{||}; j\ind{e}^{||}}_{\omega}$ of this tensor, where only parallel components of the current density vectors are correlated as already introduced in Sec. II. As it will be shown in the following sections, this bi-local current-density correlation is important to understand the excitation modes of nano plasmas. The non-diagonal components of the bi-local correlation tensor are small in comparison to the diagonal components. Beside the bi-local current density correlation function considered here, the bi-local density fluctuation correlation $\langle \delta n (\vec r), \delta n (\vec r') \rangle_{\omega}$ as well as the bi-local force correlation $\langle \vec F (\vec r), \vec F (\vec r') \rangle_{\omega}$ are useful quantities in the context of optical properties. These correlations can be evaluated from the trajectory in a similar way and are related to the bi-local current density correlation. This will be discussed in an upcoming paper.

Because of the spherical symmetry of the cluster geometry during excitation and expansion, the volume is divided into sections $\Delta V_{\vec r}$ according to $N_{r}$, $N\ind{\theta}$, $N\ind{\phi}$ equidistant intervals of spherical coordinates, i.e. the distance $r$ to the center of the cluster, the inclination angle $\theta$ as well as the azimuthal angle $\phi$, respectively. The cluster radius $R\ind{i}$ is given by the root mean square radius  of ions according to $R^2\ind{i}= 5/3 \langle r^2 \rangle$. The sections  are numbered by a single counter $a = N_{\phi} N_{\theta} \, (k - 1) + N_{\phi} \, (j - 1) + i$ with three independent counters according to the three coordinates: $i = 1 .. N_{\phi}$, $j = 1 .. N_{\theta}$ and $k = 1 .. N_{r}$. With respect to \Eq{mdcorrt} the bi-local correlation matrix $D_{a;a'} (t) = \brac{j\ind{e}^{||} (\vec r_a, 0); j\ind{e}^{||} (\vec r_{a'}, t)}$ for the spatially resolved cluster and its Laplace transform $D_{a;a'} (\omega) = \int_0^{\infty} {\rm d} t \, {\rm e}^{{\rm i} \omega t} \, D_{a;a'} (t)$ have been calculated.

The total current density ACF can be calculated from the trajectories directly. Please note, that it can be also calculated from the bi-local current density correlation matrix
\be
\mean{j\ind{e}^{||}; j\ind{e}^{||}}_{\omega} = \frac{1}{V\ind{cl}^2} \sum_{a, a'} D_{a, a'} (\omega) \Delta V_{i,j,k} \Delta V_{i',j',k'},
\label{ImpulsAKF}
\ee
using the cluster volume $V\ind{cl} = \tfrac{4 \pi}{3} R\ind{i}^3$ and the individual cell volumes
\bea
\Delta V_{i,j,k} & = & \int_{2 \pi (i - 1) / N\ind{\phi}}^{2 \pi i / N\ind{\phi}} \, {\rm d} \phi \,  \int_{\pi (j - 1) / N\ind{\theta}}^{\pi j / N\ind{\theta}} \, {\rm d} \theta \, \int_{R\ind{i} (k - 1) / N_{r}}^{R\ind{i} k / N_{r}} \, {\rm d} r \, r^2 \sin \theta \nn\\
& = & \frac{2 \pi}{3 N\ind{\phi}} \brac{\frac{R\ind{i}}{N_{r}}}^3 \brac{3 k^2 - 3 k + 1} \brac{\cos \Brac{\frac{\pi}{N\ind{\theta}} (j - 1)} - \cos \Brac{\frac{\pi}{N\ind{\theta}} j}}.
\eea
The consistency of these expressions has been checked throughout our explicit calculations.

\section{From bi-local correlation function to excitation modes}

In the following, we discuss calculations for the current-density ACF, \Eq{ImpulsAKF}, and the bi-local current-density correlation spectrum $D_{a,a'} (\omega)$. Exemplarily, we present results for the Na$\ind{309}$ cluster at electron temperature $T\ind{e} = 1$ eV, cluster charge $Z = 16$ and ionic density $n\ind{i} = 2.80 \cdot 10^{22}$ cm$^{-3}$. The electrons form a nano plasma with nonideality parameter $\Gamma = 6.964$ and degeneracy parameter $\Theta = 0.664$. Starting with a solid density cluster, these are typical parameters obtained directly after the interaction with a short pulse laser of 100 fs duration and intensity of $I = 5 \cdot 10^{11}$ Wcm$^{-2}$. Calculations of other cluster sizes will be presented in the following sections.

\begin{figure}[h]
\begin{center}
\includegraphics*[width=0.7\textwidth]{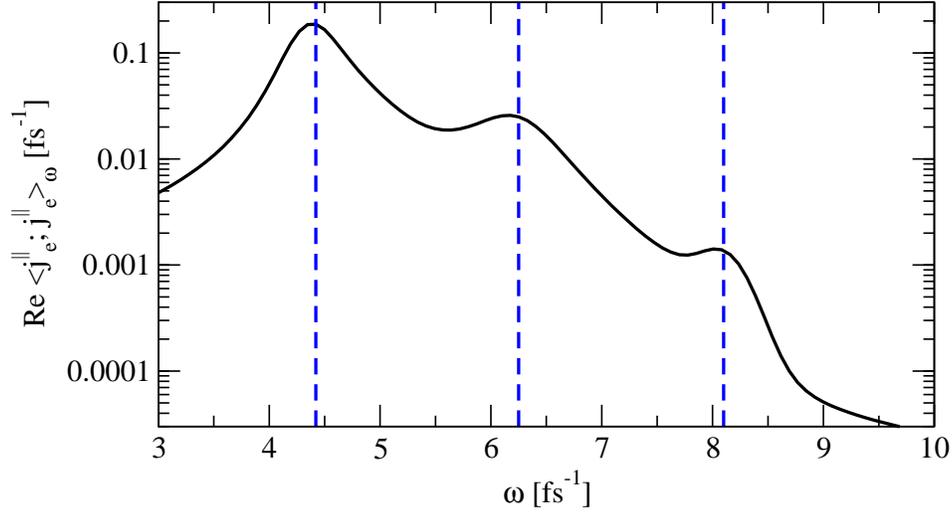}
\end{center}
\caption{(Color online) Frequency spectrum of the real part of the current-density ACF of a Na$\ind{309}$ cluster at electron temperature $T\ind{e} = 1$ eV, cluster charge $Z = 16$ and ionic density $n\ind{i} = 2.80 \cdot 10^{22}$ cm$^{-3}$.}
\label{ktot3D}
\end{figure}

The real part of the total current-density ACF ${\rm Re} \mean{j\ind{e}^{||}; j\ind{e}^{||}}_{\omega}$ is shown in \Fig{ktot3D}. Three maxima are obtained. This feature differs from the bulk behavior and is interpreted as different resonances of the electron system. To investigate the origin of the different maxima as collective excitations of the nano plasma, the bi-local current-density correlation matrix was calculated as well.

The following spatial symmetries in the matrix $D_{a;a'} (\omega)$ were found
\bea
D_{i,j,k;\ i',j',k'} (\omega) & = & D_{\left|i - i'\right| + 1,j,k;\ \left|i - i'\right| + 1,j',k'} (\omega),\label{symm1}\\
D_{i,j,k;\ i',j',k'} (\omega) & = & D_{i,N\ind{\theta} - j + 1,k;\ i',N\ind{\theta} - j' + 1,k'} (\omega),\label{symm2}\\
D_{i,j,k;\ i',j',k'} (\omega) & = & D_{i,j,k';\ i',j',k} (\omega).\label{symm3}
\eea
In our case, the $N\ind{sec}^2$ elements of the full matrix can be reduced to $N\ind{ind} = (N_{r} \, N_\theta + 1) \, N_{r} \, N_\theta \brac{N\ind{\phi} + N\ind{\phi} {\rm mod} 2} / 4$ independent elements due to the symmetries \Eq{symm1} - \Eq{symm3}, thus improving statistics via averaging equal elements.

Because of the different size of section volumes in spherical coordinates there are large variations in the mean number of particles in a section. Provided that we have $N_e = 50-1000$ electrons and $N\ind{sec} = N_\phi N_\theta N_r=128$ sections the average number of particles in some sections can be even smaller than unity. In this case, the local current density \Eq{momdens} is affected by strong fluctuations due to  the discrete number of particles. This problem is reduced, when you consider the current $\vec J\ind{e} (\vec r, t) = j\ind{e}(\vec r, t) \Delta V_{\vec r}$ as the contribution of smaller cells will be damped. Therefore, we used the non-normalized  form of the correlation function for further analysis
\be
K_{a,a'} (\omega) = D_{a;a'} (\omega) \Delta V_{i,j,k} \Delta V_{i',j',k'},
\label{ImpulsKKF}
\ee
for which the following eigenproblem was solved,
\be
\sum_{a'} \, {\rm Re}\, K_{a,a'} (\omega) \Psi_{\mu,a'} (\omega) = K_{\mu} (\omega) \Psi_{\mu,a} (\omega).
\label{EWP}
\ee
Thus, the matrix is decomposed into eigenvectors $\Psi_{\mu, a} (\omega) = \Psi_{\mu,i,j,k} (\omega)$ as well as their eigenvalues $K_{\mu} (\omega)$ at each frequency. The eigenvectors represent the spatial structure of the mode ($\Psi_{\mu,i,j,k} (\omega) \to \Psi_{\mu} (\vec r, \omega)$). The orthonormality condition
\be
\int {\rm d}^3 \vec r \, \Psi_\mu (\vec r, \omega) \, \Psi_{\mu'} (\vec r, \omega) = \delta_{\mu, \mu'}.
\ee
holds.

\begin{figure}[h]
\begin{center}
\includegraphics*[width=0.7\textwidth]{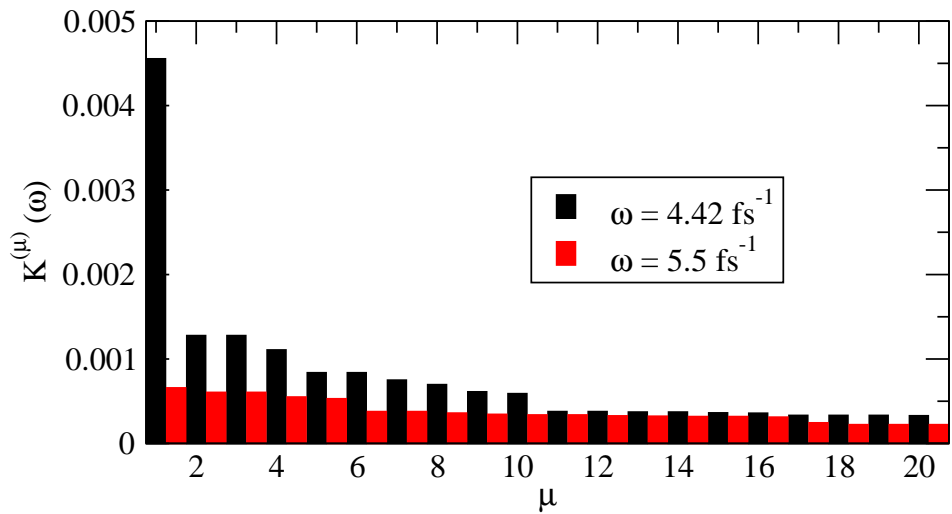}
\end{center}
\caption{(Color online) Eigenvalues of the Na$\ind{309}$ cluster sorted by size for the resonant case at $\omega = 4.42$ fs$^{-1}$ (black) and a non-resonant case at $\omega = 5.50$ fs$^{-1}$ (shaded, red online), for parameters as in \Fig{ktot3D}.}
\label{eigen}
\end{figure}

For two selected frequencies, the 10 strongest eigenvalues of the Na$\ind{309}$ cluster are shown in \Fig{eigen}. At $\omega = 4.42$ fs$^{-1}$ (black), a resonance frequency was found with one outstanding, leading eigenvalue. The second and third largest eigenvalue are of same strength, which suggests degeneracy due to the symmetry of the correlation matrix. At off-resonant frequencies, i.e. at $\omega = 5.50$ fs$^{-1}$ (shaded, red online), all eigenvalues are of the same order of magnitude.

\begin{figure}[h]
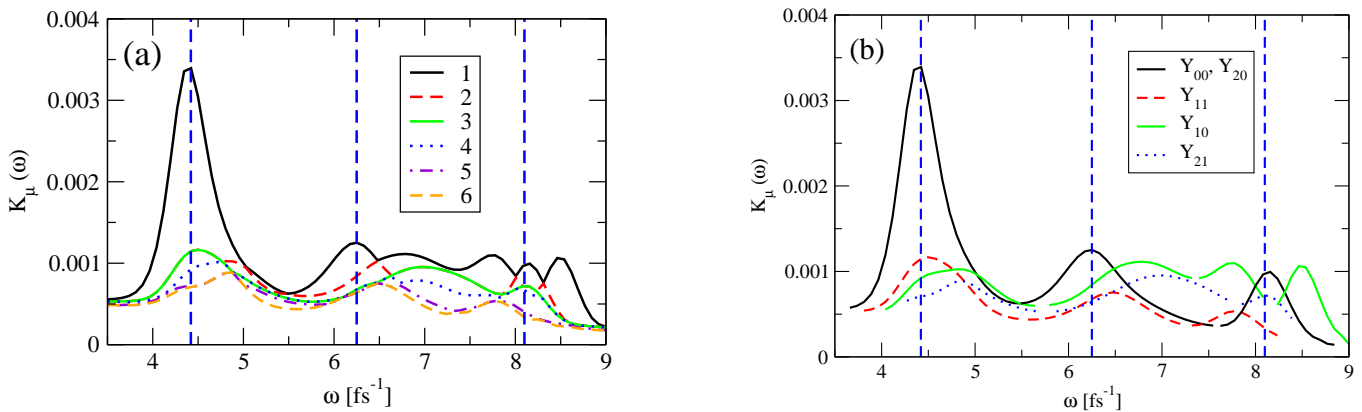

\begin{minipage}[h]{\textwidth}
\begin{minipage}[h]{0.45\textwidth}
\begin{center}
\includegraphics*[width=\textwidth]{4_eval_r.eps}
\end{center}
\end{minipage}\hfill
\begin{minipage}[h]{0.45\textwidth}
\begin{center}
\includegraphics*[width=\textwidth]{5_evalres309_r.eps}
\end{center}
\end{minipage}
\end{minipage}
\caption{(Color online) Spectrum of the 6 highest eigenvalues $K_\mu (\omega)$ of the Na$\ind{309}$ cluster for same parameters as in \Fig{ktot3D} (a). Eigenvalues of the same cluster selected in terms of the corresponding spherical harmonics $Y_{l,m} (\theta, \phi)$ (b).}
\label{evalres}
\end{figure}

In \Fig{evalres} (a) the strongest eigenvalues $K\ind{\mu} (\omega)$ of the Na$\ind{309}$ cluster are shown in dependence of frequency. They are colored according to their strength and numbered ascending with descending strength. In the shown frequency range, modes $K\ind{\mu} (\omega)$ with well defined maxima are found. The spatial oscillation structure can be identified by analyzing the eigenvectors.

In \Fig{evalres} (b), the spectra of eigenvalues are sorted in an alternative way, according to the  spatial  structure of the eigenvector which is obtained over the whole frequency range. Overall, the black solid mode is the strongest. Its resonance frequencies are also found in the total current-density ACF (indicated via vertical blue dashed lines) and are therefore of particular interest. Resonances in the total current-density ACF, shown in \Fig{ktot3D}, are only possible in the case of non-zero total current, which is caused by a dipole-like oscillation. Thus, resonances which are seen in the total current-density ACF are oscillation modes with a dipole moment. Other resonance structures, for example, are breathing modes that have no dipole moment. After characterization of the resonance structures, the dipole-like resonances will be investigated in more detail.

\section{Analysis of the collective modes}

The decomposition of the locally resolved current correlation matrix into eigenvalues $K_\mu (\omega)$, as shown in \Fig{evalres}, gives a very complex set of resonance structures in comparison to the 1D case, see \cite{Raitza10}. The spatial mode structures in 1D chains were characterized by their wave number $k$. To analyze the more complicated spatial oscillation structure of 3D clusters, a spherical Fourier decomposition of the eigenvectors into the spherical Bessel function $j_l (k_{n,l} r)$ and spherical harmonics $Y_{l,m} (\theta, \phi)$ was performed according to
\be
\Psi_\mu (\vec r, \omega) = \sum_{n = 1}^{N_n} \sum_{l = 0}^{N_l} \sum_{m = -l}^l S_{n,l,m} (\omega) \, N_{n,l} j_l (k_{n,l} r) Y_{l,m} (\theta, \phi),
\ee
where $S_{n,l,m} (\omega)$ is the spherical Fourier component with ordinal numbers $n,l,m$. The normalization factor $N_{n,l}$ as well as the wave number $k_{n,l}$ are chosen in the way that the eigenvector has a root at the cluster surface.

In \Fig{evalres} (b), the four strongest eigenvalue modes are characterized by pairs of ordinal numbers $l,m$ which determine the main angular part of the eigenvector by the spherical harmonics $Y_{l,m} (\theta, \phi)$. The leading dipole-like mode, represented via solid black lines in \Fig{evalres} (b), is characterized by the overlap of the spherical harmonic functions $Y_{0,0} (\theta, \phi)$ and $Y_{2,0} (\theta, \phi)$. For the Na$\ind{309}$ cluster, one can find three resonance frequencies which are identical to the ones found in the total current-density ACF. The latter are indicated by vertical dashed lines (blue online) in \Fig{evalres} (b).

\begin{figure}[h]
\begin{minipage}[h]{\textwidth}
\begin{minipage}[h]{0.22\textwidth}
\begin{center}
$\omega = 4.80$ fs$^{-1}$ 
\includegraphics*[width=\textwidth]{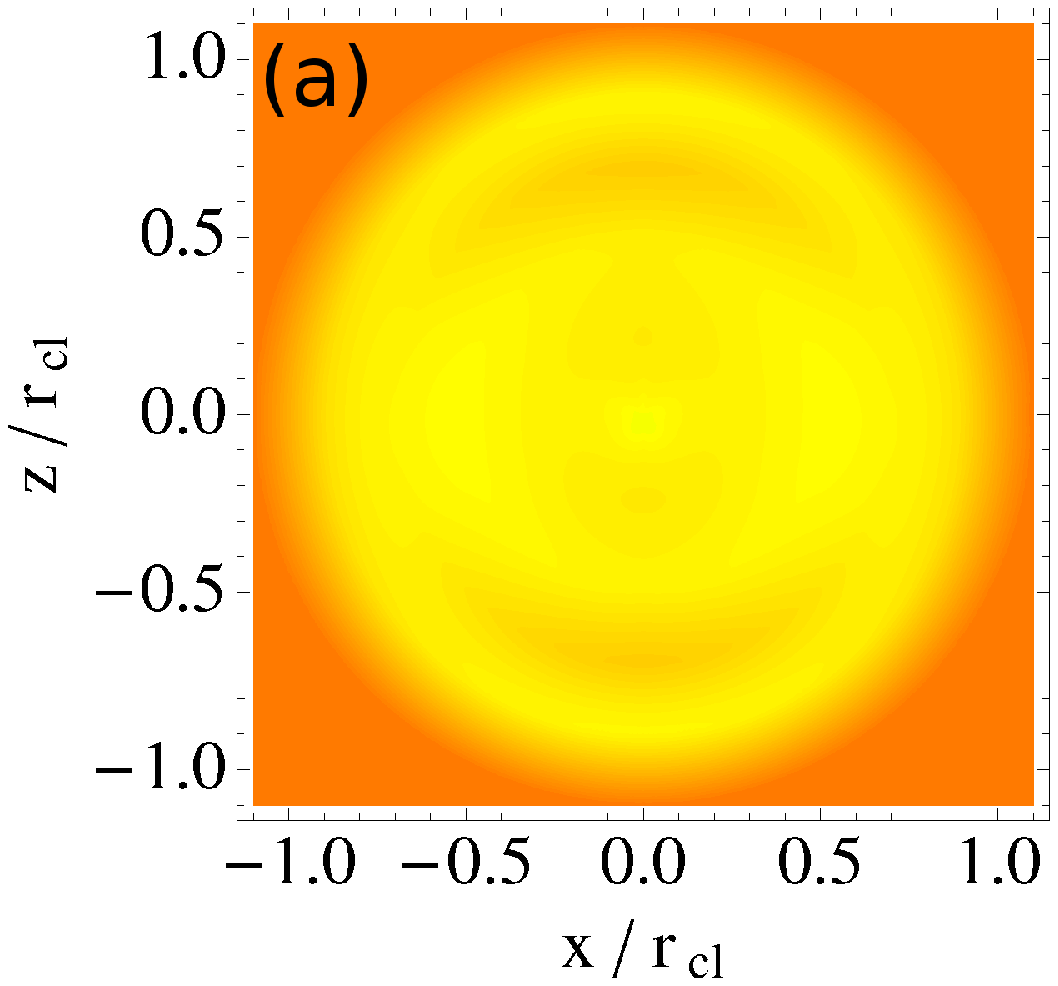}
\end{center}
\end{minipage}\hfill
\begin{minipage}[h]{0.22\textwidth}
\begin{center}
$\omega = 6.18$ fs$^{-1}$
\includegraphics*[width=\textwidth]{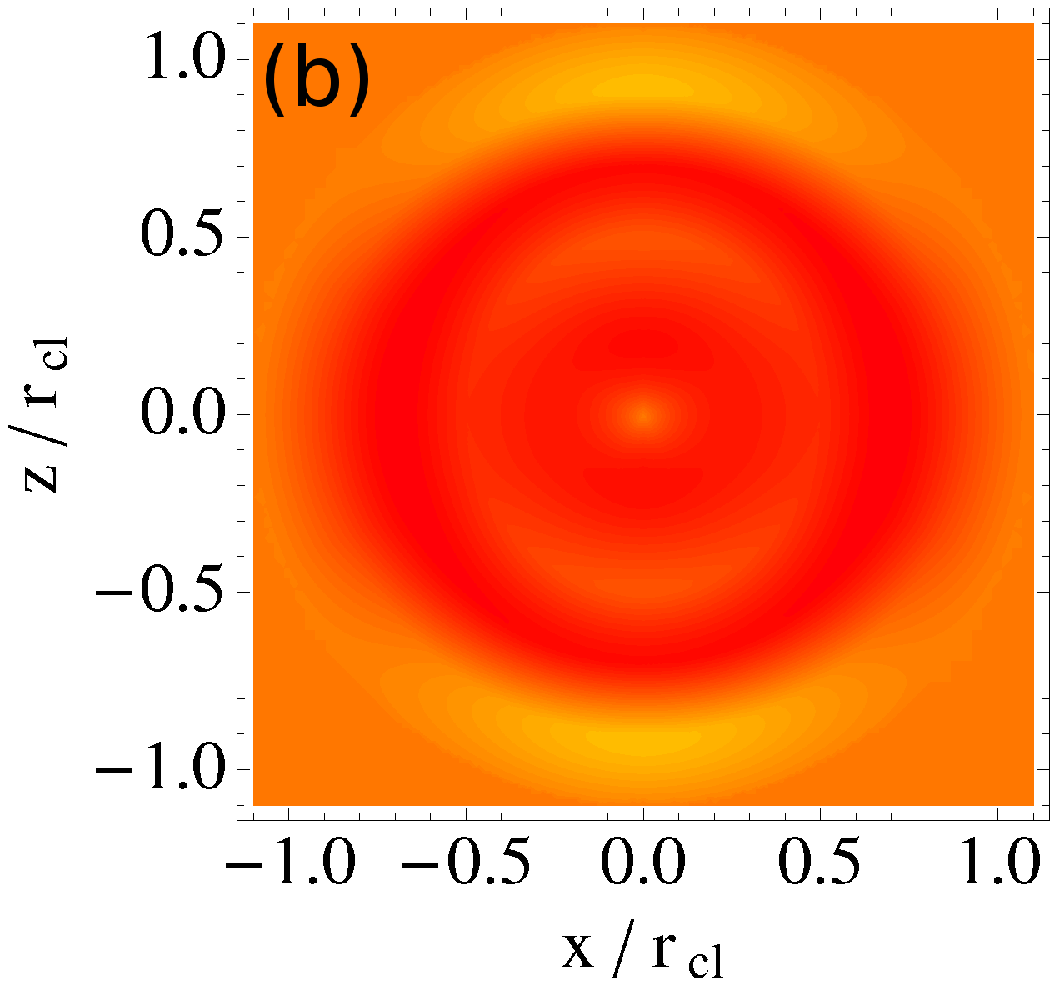}
\end{center}
\end{minipage}\hfill
\begin{minipage}[h]{0.22\textwidth}
\begin{center}
$\omega = 9.15$ fs$^{-1}$
\includegraphics*[width=\textwidth]{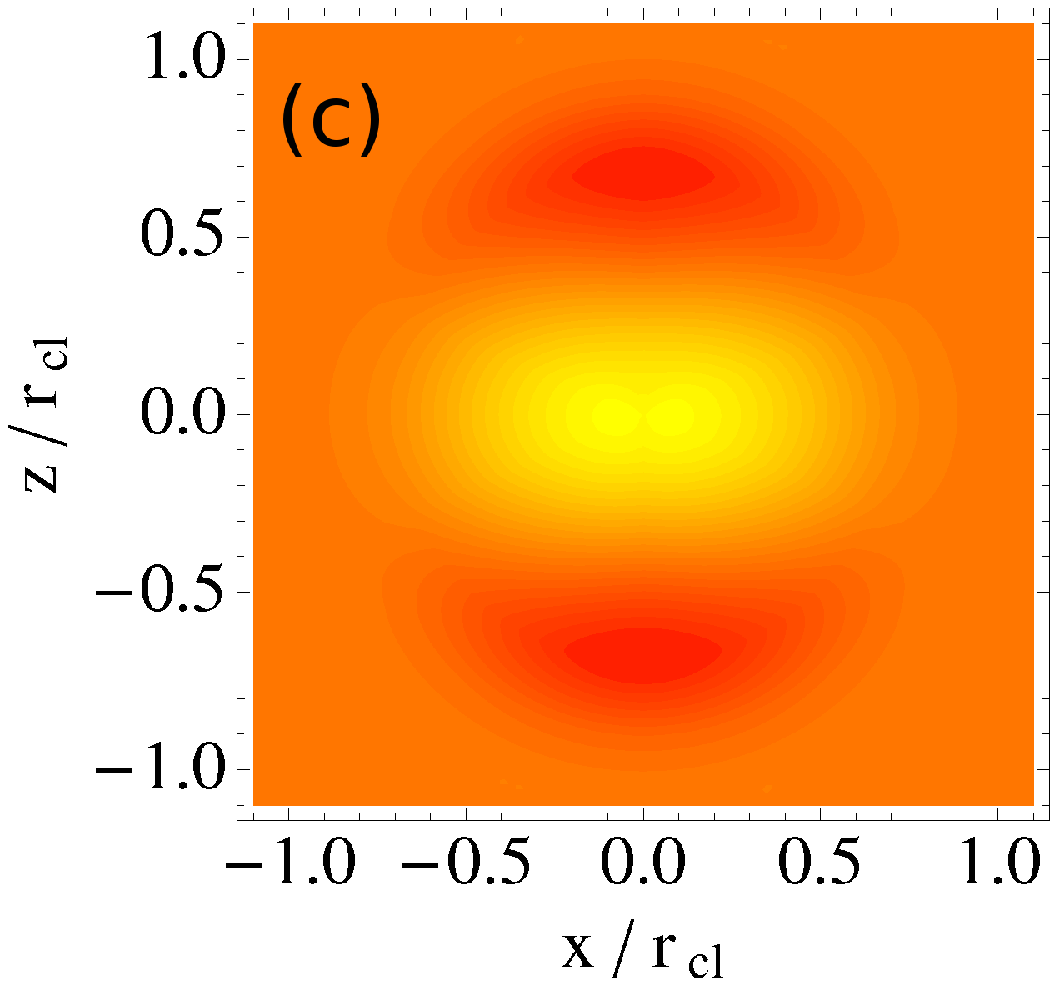}
\end{center}
\end{minipage}\hfill
\begin{minipage}[h]{0.22\textwidth}
\begin{center}
$\omega = 8.23$ fs$^{-1}$
\includegraphics*[width=\textwidth]{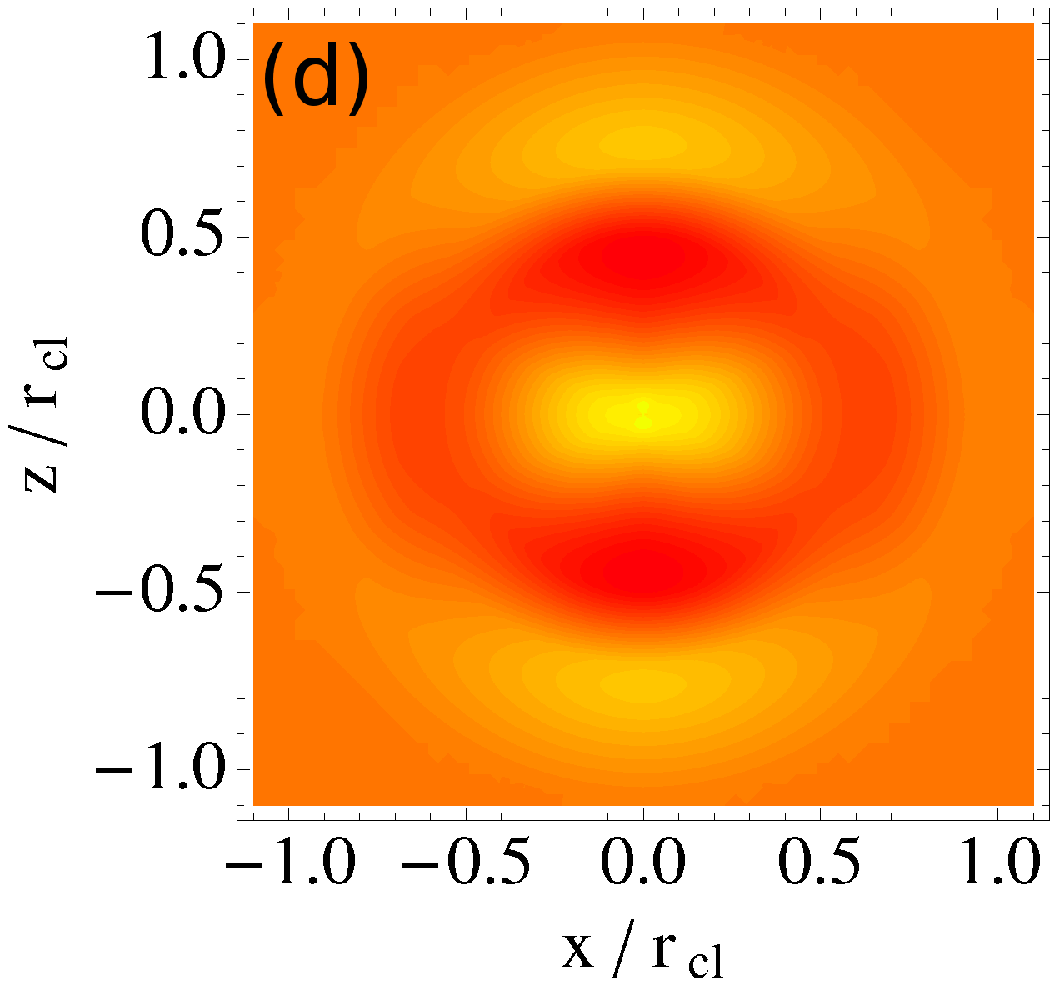}
\end{center}
\end{minipage}\hfill
\begin{minipage}[h]{0.1\textwidth}
\begin{center}
\includegraphics*[width=\textwidth]{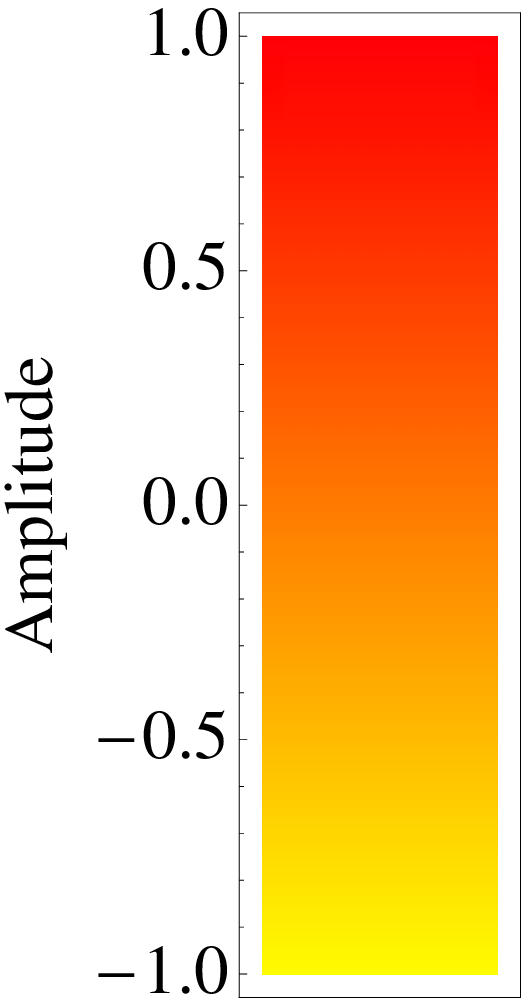}
\end{center}
\end{minipage}
\end{minipage}
\caption{(Color online) Selected eigenvectors of the dipol-like mode in the Na$\ind{1000}$ clusters for same parameters as in \Fig{ktot3D} but $Z = 19$.}
\label{evecdip}
\end{figure}

In our investigations, we looked at other cluster parameters as well and found similar behavior. Comparisons will be made in the following chapters. For further analysis of the  exication modes, we now consider a larger cluster consisting of 1000 ions. There, four pronounced dipole-like resonances were found. In \Fig{evecdip}, the spatial structures  of the current-density $j^{||}\ind{e} (\vec r) \sim \tfrac{\Psi (\vec r)}{\Delta V (\vec r)}$ is shown for the Na$\ind{1000}$ cluster at the resonance frequencies of the leading dipole-like mode. The behavior is shown in the $z-x-$plane at a fixed azimuthal angle $\phi$ on  which it does not depend.

At the resonance frequency $\omega\ind{R} = 4.80$ fs, the electrons are oscillating with a current density $j^{||}\ind{e} (r) = v (r) n\ind{e} (r)$. As shown in \Fig{evecdip} (a), all electrons of this mode are moving in the same direction and no nodes can be seen.
Assuming a constant velocity field amplitude $v = {\rm const}$, the change of the current density with distance $r$ is directly related to the density profile $n\ind{e} (r)$ of the electrons. 

The modes in \Fig{evecdip} (c) and (d) are similar to a plane wave oscillation of electrons, but trapped inside the cluster. To identify a wave number of the plane wave oscillation, a Fourier decomposition of plane waves in $z$-direction was done. A maximum at $k = 1.6$ nm$^{-1}$ and $k = 4.7$ nm$^{-1}$, respectively, is found which  identify the wavelengths of the plane wave oscillations. Only in the large cluster with 1000 ions, a plane wave oscillation with higher wavenumber was found. All other modes can be seen in smaller clusters as well. The resonance structure in \Fig{evecdip} (b) looks like a mix of the first and the third resonance structure.

We want to point out one further feature of the mode spectra in \Fig{evalres} (b). The dashed red line represents in fact two resonance structures with exactly the same eigenvalues at all frequencies. The eigenvectors are orthogonal since they are characterized by the same spherical harmonic function $Y\ind{1,1} (\theta, \phi)$ but have a phase shift in $\phi$-direction: $Y^{(1)}\ind{1,1} (\theta, \phi) = Y^{(2)}\ind{1,1} (\theta, \phi + \tfrac{\pi}{2})$. Further degenerations are obtained for weaker eigenvalue modes as well.

All eigenvectors $\Psi_\mu (\vec r, \omega)$ are decomposed into a superposition of spherical Bessel functions $j_l (k_{n,l} r)$ with a set of ordinal numbers $n$. No leading ordinal number $n$ was found, which characterizes the spatial resonance structure in $r$ direction.

\subsection{Resonance frequency of the rigid oscillation}

The total current density ACF shown in \Fig{ktot3D} as well as the leading eigenvalue mode in \Fig{evalres} (right) show the strongest resonance at the frequency $\omega\ind{R} \approx 4.42$ fs$^{-1}$. This resonance belongs to the dipole-like mode with the eigenvector shown in \Fig{evecdip} on the left hand side. We will now analyze this collective excitation mode in terms of a rigid oscillation.

The electrons with density profile $n\ind{e} (\vec r)$ are assumed to move nearly rigidly in the external potential $V\ind{ext,ei} (\vec r)$ due to the fixed ions. The potential energy of the electrons due to a small shift with respect to the ions reads
\be
U\ind{e} (z) = \int {\rm d}^3 \vec r \, n\ind{e} (\vec r) \, V\ind{ext, ei} (\vec r - z \vec e_z).\label{epot}
\ee
The change of the potential energy $U\ind{e} (z)$ in $z$ direction is due to the restoring force on the electron profile. In harmonic approximation of the equation of motion, the resonance frequency is identified as
\be
m\ind{e} N\ind{e} \omega\ind{R}^2 = - \left. \dderive{U\ind{e} (z)}{z}\right|_{z = 0}.\label{eom}
\ee
For small rigid shifts $z \rightarrow 0$, assuming radially dependent electron density profiles and external potentials in \Eq{epot} the integration over the angular dependence of the potential energy calculation can be executed. The resonance frequency \Eq{eom} is then given according to
\be
\omega\ind{R}^2 = \frac{4 \pi}{3 m\ind{e} N\ind{e}} \int_0^{\infty} {\rm d} r \, n\ind{e} (r) \, r^2 \, \brac{V\ind{ext, ei}'' (r) + 2 \frac{V\ind{ext, ei}' (r)}{r}}.\label{omegaR}
\ee

As a first example for a density profile, we assume a homogeneously charged ion sphere with radius $R\ind{i}  = (3 N\ind{i} / (4 \pi n\ind{i}) )^{1/3}$ and an electron sphere with radius $R\ind{e}  = (3 N\ind{e} / (4 \pi n\ind{e}) )^{1/3}$. The densities of the electron and ion spheres is equal ($n\ind{e} = n\ind{i}$). Therefore, the difference of ion and electron radius is determined by the cluster charge, basically the difference of the simulated electron number $N\ind{e}$ and ion number $N\ind{i}$. Thus, in the case of positively charged clusters, as discussed here, the electron sphere radius is smaller than the ion radius ($R\ind{e} < R\ind{i}$). The error function potential \Eq{erf} was taken as electron-ion-interaction potential for the calculation of the resonance frequency, as it was used for the MD simulation as well. The resonance frequency than reads
\bea
\omega\ind{R}^2 (R\ind{i}, R\ind{e}) & = & \omega\ind{Mie}^2 \left[\frac{R\ind{i}^3 + R\ind{e}^3}{2 R\ind{e}^3} {\rm erf} \brac{\frac{R\ind{i} + R\ind{e}}{\lambda}} - \frac{R\ind{i}^3 - R\ind{e}^3}{2 R\ind{e}^3} {\rm erf} \brac{\frac{R\ind{i} - R\ind{e}}{\lambda}} +\right.\nn\\
&&\left. \frac{{\rm e}^{-\frac{R\ind{i}^2 + R\ind{e}^2}{\lambda^2}}}{\sqrt{\pi} R\ind{e}^3} \brac{\Brac{\frac{\lambda^3}{2} - \lambda \brac{R\ind{i}^2 + R\ind{e}^2}} {\rm sinh} \brac{\frac{2 R\ind{i} R\ind{e}}{\lambda^2}} - \lambda R\ind{i} R\ind{e} {\rm cosh} \brac{\frac{2 R\ind{i} R\ind{e}}{\lambda^2}}}\right].\label{oRerf}
\eea
In the limit of large clusters with high number of ions the resonance frequency equals the Mie frequency,\linebreak $\lim_{R\ind{i} \to \infty} \omega\ind{R} (R\ind{i}) = \omega\ind{Mie}$. Assuming only a weak charged cluster, the sphere radii have nearly the same size ($R\ind{e} \rightarrow R\ind{i}$) and the system is nearly neutral. The limit for small clusters, down to just one atom, depends strongly on the pseudopotential. In our case, the resonance frequency $\lim_{N\ind{i} \to 1} \omega\ind{R} (N\ind{i}) = e^2/4 \pi \ep\ind{0} \cdot  4/(3 \sqrt{\pi} \lambda^3 m\ind{e})$ is due to the oscillation of a single electron in the ionic error-function pseudo-potential \Eq{erf}.

\begin{figure}[h]
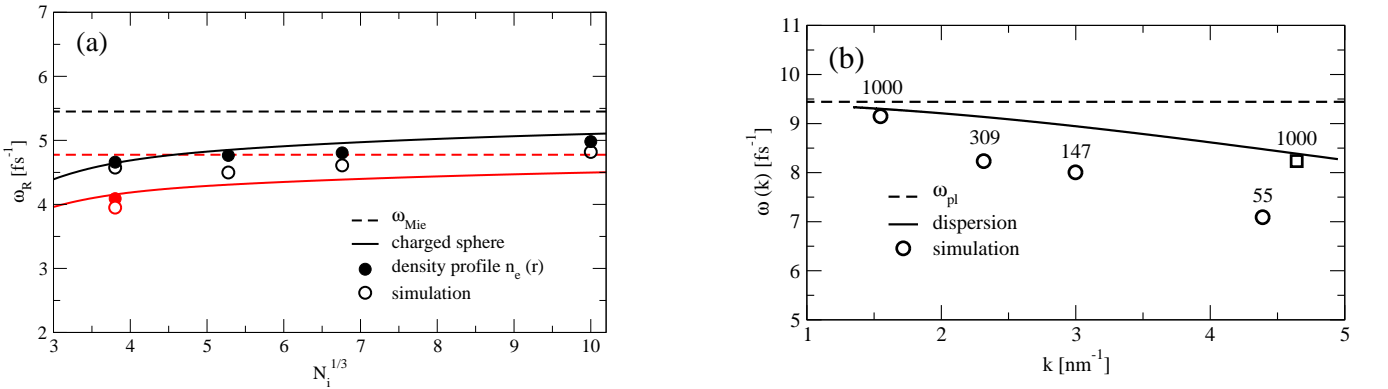

\begin{minipage}[h]{\textwidth}
\begin{minipage}[h]{0.45\textwidth}
\begin{center}
\includegraphics*[width=\textwidth]{11_omegaRn_r.eps}
\end{center}
\end{minipage}\hfill
\begin{minipage}[h]{0.45\textwidth}
\begin{center}
\includegraphics*[width=\textwidth]{12_omegak_r.eps}
\end{center}
\end{minipage}
\end{minipage}
\caption{(Color online) Cluster size dependent resonance frequency $\omega\ind{R} (R\ind{i})$ (a). Simulation results (empty circles) and analytical calculations (solid lines) using \Eq{oRerf} are shown for $n\ind{i} = 2.15 \cdot 10^{22}$ cm$^{-3}$ (shaded, red online) and $n\ind{i} = 2.80 \cdot 10^{22}$ cm$^{-3}$ (black). The Mie frequencies are  given as dashed lines. Numerical calculations using \Eq{omegaR} are presented (full dots). Dispersion \Eq{omk} of a plane wave in a homogeneously charged sphere (solid line) for $n\ind{e} = 2.80 \cdot 10^{22}$ cm$^{-3}$ as well as simulation results (empty symbols) and bulk plasmon frequency (b).}
\label{omegaRfig}
\end{figure}

In \Fig{omegaRfig} (a), the resonance frequency $\omega\ind{R}$ of the dipole-like mode is shown in dependence on the size of the ion sphere. Results from MD simulations (empty circles) for  Na$\ind{55}$, Na$\ind{309}$ and Na$\ind{1000}$ cluster at $n\ind{i} = 2.80 \cdot 10^{22}$ cm$^{-3}$  as well as the Na$\ind{55}$ cluster at $n\ind{i} = 2.15 \cdot 10^{22}$ cm$^{-3}$ are shown. The resonance frequencies  have been calculated using \Eq{oRerf} for ion densities of $n\ind{i} = 2.15 \cdot 10^{22}$ cm$^{-3}$ (solid shaded line, red online) and $n\ind{i} = 2.80 \cdot 10^{22}$ cm$^{-3}$ (solid black line). The limits of large clusters, the Mie frequency $\omega^2\ind{Mie} = e^2 n\ind{i}/(3 \ep\ind{0} m\ind{e})$, are given as dotted lines colored according to the two densities.

Additionally, the electron density profile $n\ind{e} (r)$ was deducted from MD simulations for all cluster sizes and used to derive the resonance frequency $\omega\ind{R}$ solving \Eq{omegaR} numerically. As a result (full circles in \Fig{omegaRfig} (a)), the resonance frequency of the dipole-like mode is obtained with a deviation to the direct simulation results of less than 5\%. Using homogeneously charged ion and electron spheres leads to reasonable agreement in the limits of large clusters as well as for small clusters. Taking the spatial structure of the density profile into account, there is good agreement with the direct  simulation results   in the intermediate cluster size regime as well.

\subsection{Dispersion of the plane wave mode}

While the Mie-like resonance, discussed in the previous subsection, is almost spherically symmetric, we obtain an increasing plane wave character of the dipole-like mode with increasing frequency. The third resonance frequency of the total current density ACF for the Na$\ind{309}$ cluster at $\omega\ind{R} = 8.14$ fs$^{-1}$, see \Fig{ktot3D}, is mainly caused by a plane wave like eigenvector, which is similar to the eigenvector of the Na$\ind{1000}$ cluster, shown in \Fig{evecdip} (right). This mode is discussed by Kresin {\it et al.} \cite{Kresin09} as compressional volume plasmon. Here, oscillations of electrons in opposite directions must be taken into account for the analytical calculation of the resonance frequency. We assume homogeneously charged spheres for the electrons with radius $R\ind{e}$ and for the ions with radius $R\ind{i}$ as it was already discussed in the previous subsection. The electron motion is treated as a hydrodynamical liquid using the Euler equation
\be
\derive{\vec j (\vec r,t)}{t} = -{\rm div}\, \Brac{\vec j (\vec r, t) \otimes \vec v(\vec r, t)} - \frac{1}{m\ind{e}} {\rm grad}\, p (\vec r, t) - \frac{n\ind{e} (\vec r, t)}{m\ind{e}} {\rm grad}\, V\ind{ext} (\vec r, t),\label{euler}
\ee
where $\vec j (\vec r,t) = n\ind{e} (\vec r,t) \vec v(\vec r, t)$ is the spatially resolved current density of the electrons, $p (\vec r, t)$ is the pressure of the electron gas and $V\ind{ext} = V\ind{ext,ei} + V\ind{ext,ee}$ is the external potential, composed of contributions from the electrons and ions. Using the following ansatz
\bea
j_{z} (\vec r, t) & = & \delta j \vec e_z {\rm e}^{{\rm i} (k z - \omega t)},\\
v_{z} (\vec r, t) & = & \delta v \vec e_z {\rm e}^{{\rm i} (k z - \omega t)},\\
n\ind{e} (\vec r, t) & = & n\ind{e,0} (r) + \delta n\ind{e} {\rm e}^{{\rm i} (k z - \omega t)},\\
V\ind{ext} (\vec r, t) & = & V\ind{ext, 0} + \delta V\ind{ext} (\vec r),
\eea
we consider small perturbations in $z$-direction restricting ourselves to longitudinal effects. One is able to linearize the Euler equation. The system is assumed to be in LTE, described by the quantities $n\ind{e,0} (r)$, $\vec j\ind{0} (\vec r) = 0$, $\vec v\ind{0} (\vec r) = 0$ as well as $V\ind{ext, 0} (r)$. Electrons are moving in the external field of ions and in the mean field of electrons. The external potential is
\bea
V\ind{ext} (\vec r, t) & = & \int {\rm d}^3 \vec r\ind{1} \brac{n\ind{i} (r\ind{1}) - \frac{n\ind{e,0} (r\ind{1})}{2}} V\ind{e,i} (\vec r\ind{1} - \vec r) + \frac{1}{2} \int {\rm d}^3 \vec r\ind{1} \delta n\ind{e} (r\ind{1}, t) V\ind{e,e} (\vec r\ind{1} - \vec r),\nn\\
V\ind{ext} (\vec r, t) & = & V\ind{ext, 0} (r) + \delta V\ind{ext} (\vec r, t).\label{Vextpert}
\eea
The external potential has a equilibrium part and a perturbative part $\delta V\ind{ext} (\vec r, t)$, which is mainly dependent on the linear density perturbation $\delta n\ind{e} (\vec r, t)$.

Assuming Boltzmann distribution we express the ideal gas pressure of the electrons $p (\vec r, t)$ via the electron density.  Using the equation of continuity, one is able to express the Euler equation in terms of linear perturbations of the density. Thus, the equilibrium part of the external potential compensates the pressure term on the right hand side of \Eq{euler}. Restricting ourselves to linear perturbations of the Euler equation, only the third term on the right hand side of \Eq{euler} remains, which is connected to the external potential. Finally, all terms of the Euler equation \Eq{euler} are lead back to a linear density fluctuation $\delta n\ind{e} (\vec r, t)$. Thus, one ends up with
\bea
\omega^2 (k) & = & \frac{\omega\ind{pl}}{4} {\rm e}^{-\frac{k}{4} \brac{k \lambda^2 + 4 {\rm i} R\ind{e}}}\nn\\
&& \brac{{\rm i} {\rm e}^{{\rm i} k R\ind{e}} \Brac{{\rm erfi} \brac{\frac{k \lambda^2 - 2 {\rm i} k R\ind{e}}{2 \lambda}} - {\rm erfi} \brac{\frac{k \lambda^2 + 2 {\rm i} k R\ind{e}}{2 \lambda}}} - {\rm e}^{\frac{k^2 \lambda^2}{4}} \Brac{1 + {\rm e}^{2 {\rm i} k R\ind{e}}} {\rm erf} \brac{\frac{R\ind{e}}{\lambda}}}.\label{omk}
\eea
This relation leads to real valued solutions for the resonance frequencies for standing waves with $k_n = n \pi / R\ind{e}$ only and scales with the plasma frequency $\omega\ind{pl}$. In the limit $k \to 0$ we find  $\omega (0) = \omega\ind{pl}$ which coincides with the bulk limit.

From the eigenvector of the plane wave mode, one can derive the wave number $k = \pi / R\ind{e}$, which corresponds to $n = 1$. This means the dispersion of the plane wave mode is determined by the radius of the electron cloud. Results for this case are shown in \Fig{omegaRfig} (b) for different cluster sizes and are compared with the simulation data. For the cluster with 1000 ions a plane wave mode with $k = 3 \pi / R\ind{e}$ and $n = 3$ was found as well. In \Fig{omegaRfig} (b), it is marked with an empty square. Its spatial structure is shown in \Fig{evecdip} (d). The simulation data for the Na$\ind{1000}$ cluster fit the dispersion \Eq{omk} as well. Deviations of the plane wave resonance for smaller clusters are caused by to the radial dependence of the electron density profile.

\section{Conclusion}

We have investigated collective excitation modes of a nano plasma in highly excited metal clusters. The collective excitation of electrons inside the cluster are obtained from bi-local current density correlation functions by solving the eigenvalue problem of the current-density correlation matrix. Using RMD simulations, the local current density $\vec j (\vec r, t)$ for excited clusters of 55 up to 1000 ions with densities of $n\ind{i} = 2.15 \cdot 10^{22}$ cm$^{-3}$ as well as $2.80 \cdot 10^{22}$ cm$^{-3}$ and temperatures of $T\ind{e} = 1$ eV have been investigated. Pseudo-potentials of sodium were used to calculate the electron dynamics without consideration of degeneration effects any further. For the analysis of electron dynamics at lower temperatures, the inclusion of quantum effects for the calculation of the local current density $\vec j (\vec r, t)$ of cluster electrons is an open question at this point. It would be useful to go beyond present classical description to discuss for example cold, non-excited clusters.

The spectrum of dipole-like modes was investigated  in more detail. Using analytical calculations, it was possible to relate the position of resonance modes in the frequency domain to their spatial mode structure. Results for the cluster size dependence of the resonance frequency have been shown. A smooth transition to the bulk behavior has been obtained. The analysis of further resonance frequencies and also other modes including breathing modes would be desirable. The width of mode resonances and the role of collision-less damping effects as well as the collision frequency need to be investigated in the future. The systematic change of the collision frequency with cluster size up to the bulk limit remains an interesting field. 

From RMD simulations, different collective excitations have been found in nano plasmas, including dipole-like and breathing modes. These collective excitations will influence the scattering and absorption properties of clusters, see \cite{Kresin09}. Collective effects of electron motion play a role when analyzing ultraviolet (UPS) or x-ray photo-electron spectroscopy (XPS) experiments, as  has been pointed out by Andersson {\it et al.} \cite{Andersson11}. It is a challenge to experimentalists to confirm the occurence of different collective excitations in nano plasmas.

\section*{Acknowledgement}

We would like to acknowledge financial support of the SFB 652 which is funded by the DFG. I.~M. acknowledges support by the Programs of Fundamental Research of the Presidium of RAS Nos. 2, 13 and 14 and the Grant of President of Russian Federation No. MK-64941.2010.8. We like to thank Eric Suraud for fruitful discussions.

\end{document}